\documentclass[conference]{IEEEtran}
\usepackage{amsmath}
\usepackage{amsfonts}
\usepackage{amssymb}
\usepackage{graphicx}
\graphicspath{{figures/}}
\usepackage[noadjust]{cite}
\usepackage{subfigure}
\usepackage{float}
\usepackage{algorithm}
\usepackage{algorithmic}
\usepackage[table]{xcolor}
\usepackage{multirow}
\begin{document}
\title{Adaptive Spectrum Sharing of LTE Co-existing with WLAN in Unlicensed Frequency Bands}

\author{Minyao Xing, ~Yuexing Peng, ~Teng Xia, ~Hang Long, ~Kan Zheng\\
Wireless Signal Processing \& Network Lab\\
Key Laboratory of
Universal Wireless Communication, Ministry of Education \\
Beijing University of Posts \& Telecommunications, Beijing, China\\
Email: vivian20@bupt.edu.cn\\ }
\maketitle

\begin{abstract}
%\boldmath
With the increase of wireless communication demands, licensed spectrum for long term evolution (LTE) is no longer enough. The research effort has focused on implementing LTE to unlicensed frequency bands in recent years, which unavoidably brings the problem of LTE co-existence with other existing systems on the same band. This paper proposes an adaptive co-existence mechanism for LTE and wireless local area networks (WLAN) to enable a significant system performance of WLAN while LTE does not lose much as well. LTE realizes the co-existence by allocating time resources dynamically according to the traffic load of WLAN system.
\end{abstract}


\section{Introduction}
\label{sec:intro}
In recent years, long term evolution (LTE) standardized by the 3rd Generation Partnership Project (3GPP) has become one of the most popular cellular technologies in the world. The increasing use of smartphones and other wireless devices has caused an exponential increase in mobile broadband data usage, which results in the so-called \emph{spectrum scarcity} problem. Since the existing spectrum is not enough for LTE system and available limited licensed spectrum can be costly to increase, implementing LTE in unlicensed frequency bands is a promising way to solve the problem mentioned before. It is inevitable for LTE to co-exist with other networks when it expands over to unlicensed spectrums. Co-existence between heterogeneous systems has been studied before, for instance, IEEE 802.15.4 and IEEE 802.11b/g \cite{1}, LTE and WLAN \cite{2}. It is obvious that the co-existence causes significant degradation on system performance. Besides, interference coordination schemes have also been studied to increase spectral efficiency in \cite{3} and \cite{4}. An effective approach to reduce the interference in underlay spectrum sharing networks with multiple users and antennas by transmitting antenna selection in decode-and-forward relaying is presented in \cite{5}. Theoretical analysis of channel sharing methods is discussed in \cite{6}, and the authors perform stochastic performance analysis of a finite-state Markov channel shared by multiple users.
\par
Several studies of LTE co-existing with WLAN in unlicensed frequency bands can be found in \cite{7}-\cite{9}. In \cite{7}, the authors realize the co-existence between LTE and WLAN by limiting LTE presence in the shared band. Similarly, in \cite{8}, a mechanism to enable co-existence between LTE and WLAN in 900 MHz is described. This mechanism utilizes a modified version of almost-blank subframes (ABS) and improves the WLAN system performance by allocating more subframes to it. However, simulation results show that only when LTE gives up more than 80\% of the subframes can WLAN reach the optimum performance. Otherwise, WLAN degrades severely. In \cite{9}, the concept of deploying LTE in the license-exempt band is presented. Also, some modifications needed to be made in LTE systems: a listen-before-talk (LBT) scheme and RTS/CTS protocol. However, communication efficiency will decrease because LTE listens to the channel before every transmission. It is necessary to enable LTE co-exist with WLAN and not to degrade their performances.
\par
In this paper, we investigate an adaptive mechanism for the co-existence of LTE and WLAN. LTE does not occupy all the subframes in this mechanism but spares subframes to WLAN system dynamically. In other words, LTE adjusts its subframe configuration periodically regarding the traffic load of WLAN system. Therefore, the two systems both can deliver significant capacity during the co-existence.
\par
This paper is organized as follows. In Section \uppercase\expandafter{\romannumeral 2}, the adaptive co-existence methodology is discussed. In Section \uppercase\expandafter{\romannumeral 3}, the co-existence mechanism of LTE and WLAN system in unlicensed frequency bands is presented. Section \uppercase\expandafter{\romannumeral 4} presents the simulation scenarios and results. Finally, conclusions are drawn in Section \uppercase\expandafter{\romannumeral 5}.
\par

\section{Adaptive Co-existence Methodology}
\label{sec:co-existence methodology}
Before discussing the adaptive co-existence methodology, we should notice that LTE differs greatly from WLAN in terms of allocating bandwidth resources when transmitting data on physical medium. The main difference between LTE and WLAN system is that LTE just reduces its transmission speed while WLAN holds its transmission in the case of channel interference. WLAN is a contention based system and is based on carrier sense multiple access with collision avoidance (CSMA/CA), that is, terminals and access points (AP) contend together for the same band.
\par
\begin{table}[htbp]
\caption{Subframe Configuration Examples of LTE System.}\label{table:1}
\centering
\begin{tabular}{|c|c|c|c|c|c|c|c|c|c|c|}
\hline &\multicolumn{10}{c|}{\bf Subframe Number} \\
\hline \bf Configurations &0 &1 &2 &3 &4 &5 &6 &7 &8 &9 \\
\hline \bf Mode 0 & T &T &T &T &T &T &T &T &T &T \\
\hline \bf Mode 1 & T &\multicolumn{1}{|>{\columncolor{red!25}}c|}{L} &T &T &T &T &\multicolumn{1}{|>{\columncolor{red!25}}c|}{L} &T &T &T \\
\hline \bf Mode 2 & T &\multicolumn{1}{|>{\columncolor{red!25}}c|}{L} &\multicolumn{1}{|>{\columncolor{red!25}}c|}{L} &T &T &T &\multicolumn{1}{|>{\columncolor{red!25}}c|}{L} &\multicolumn{1}{|>{\columncolor{red!25}}c|}{L} &T &T\\
\hline \bf Mode 3 &T &\multicolumn{1}{|>{\columncolor{red!25}}c|}{L} &\multicolumn{1}{|>{\columncolor{red!25}}c|}{L} &T &\multicolumn{1}{|>{\columncolor{red!25}}c|}{L} &\multicolumn{1}{|>{\columncolor{red!25}}c|}{L} &\multicolumn{1}{|>{\columncolor{red!25}}c|}{L} &\multicolumn{1}{|>{\columncolor{red!25}}c|}{L} &T &T\\
\hline \bf Mode 4 &T &\multicolumn{1}{|>{\columncolor{red!25}}c|}{L} &\multicolumn{1}{|>{\columncolor{red!25}}c|}{L} &\multicolumn{1}{|>{\columncolor{red!25}}c|}{L} &\multicolumn{1}{|>{\columncolor{red!25}}c|}{L} &\multicolumn{1}{|>{\columncolor{red!25}}c|}{L} &\multicolumn{1}{|>{\columncolor{red!25}}c|}{L} &\multicolumn{1}{|>{\columncolor{red!25}}c|}{L} &\multicolumn{1}{|>{\columncolor{red!25}}c|}{L} &T \\
\hline
\end{tabular}
\vspace{-15pt}
\end{table}
As a result, considering the difference in bandwidth allocation, the majority of WLAN nodes are likely to be blocked by LTE when LTE and WLAN networks are deployed in the same frequency band because LTE interference levels are often above the threshold which WLAN uses to determine whether the channel is idle. Therefore, medium access control (MAC) approach is effective to realize the heterogeneous co-existence, whose essential idea is to detect and then adjust frequency, transmit power, or time resource to avoid interferences. Then we can adjust the allocation of time resource according to the traffic load of WLAN. The time resource discussed here is the subframes that LTE occupies.
In most existing studies, LTE occupies all the subframes, which results in WLAN nodes suffering a continuous and significant interference from LTE system. This interference blocks the WLAN channel and makes most WLAN nodes stay in carrier sensing (CS) stage. With our co-existence methodology, LTE dose not occupy all the subframes, and the subframe configuration will change periodically. Also, LTE listens to the channel during mute subframes and prepares for the next adjustment of subframe configuration.
\par

Table \ref{table:1} shows examples of co-existence subframe configurations. Here T denotes a LTE transmission subframe and L denotes a mute subframe in which LTE listens to the channel. Besides, we evaluate the performances of LTE and WLAN systems when LTE works with fixed subframe configurations and compare them to our adaptive co-existence mechanism.

\section{Adaptive Spectrum Sharing Mechanism}

The main purpose of adaptive co-existence mechanism is to make LTE allocate time resource adaptively according to the traffic load of WLAN system. It is significant to introduce the clear channel assessment (CCA) mechanism into LTE system. Unlike traditional listen-before-talk mechanism, LTE only listens to the channel during mute subframes and it takes the combination of energy detection (ED) and CS. ED is a universal mechanism that can be introduced into all the systems. It can independently perceive whether the channel is occupied by comparing the detected channel energy with a ED threshold. CS can report a busy channel by sensing a signal with the known characteristics. As a result, LTE can sense whether the channel is occupied by WLAN during the mute subframes.
\par
Take $T_c$ as the subframe reallocation cycle and $\gamma$ is the ratio indicating WLAN traffic load. $\gamma$ is defined as follows:
\begin{equation}
\label{equ:1}
\gamma=\frac{N_{seize}}{N_{listen}},
\end{equation}
\par
\noindent where $N_{seize}$ is the number of mute subframes in which the channel energy LTE detected is above the ED threshold and a WLAN signal can also be sensed. $N_{listen}$ is the total number of mute subframes during $T_c$. Then LTE adjusts its subframe configuration according to the $\gamma$ calculated during cycle $T_c$. The specific mechanism algorithm is defined as follows:
\begin{algorithm}
\caption{Adaptive Co-existence Mechanism.}
\label{alg1}
\begin{algorithmic}[1]
\REQUIRE ~~ \\
The ratio $\gamma$ calculated during reallocation cycle $T_c$;\\
$\gamma \geqslant 0$ and $\gamma \leqslant 1$;\\
\ENSURE ~~ \\
LTE subframe configuration modes; \\
\IF {$\gamma \leqslant 0.08$}
\STATE LTE spares 1 subframe to WLAN;
\ELSIF {$\gamma \leqslant 0.16$}
\STATE LTE spares 2 subframes to WLAN;
\ELSIF {$\gamma \leqslant 0.24$}
\STATE LTE spares 3 subframes to WLAN;
\STATE ...
\ELSIF {$\gamma \leqslant 0.86$}
\STATE LTE spares 8 subframes to WLAN;
\ELSIF {$\gamma \leqslant 0.94$}
\STATE LTE spares 8 subframes to WLAN;
\ELSE
\STATE LTE spares 9 subframes to WLAN;
\ENDIF
\end{algorithmic}
\end{algorithm}
\par
There are 13 subframe configuration modes in our adaptive co-existence mechanism since the smaller the step size of LTE reallocation is the better the system performances. It is reasonable that the heavier of WLAN traffic load is the bigger value of $\gamma$ is, and LTE will spare more subframes to WLAN in the next cycle. As a result, LTE adapts its subframe configuration according to the traffic load of WLAN and both systems can make full use of the time resource.

\section{Simulation Results}
In this section, we evaluate the performance of the proposed mechanism, and compare it with the co-existence schemes which employ fixed subframe configurations.

\subsection{Simulation Scenarios}
\begin{figure}[htbp]
\centering
\includegraphics[width=6cm]{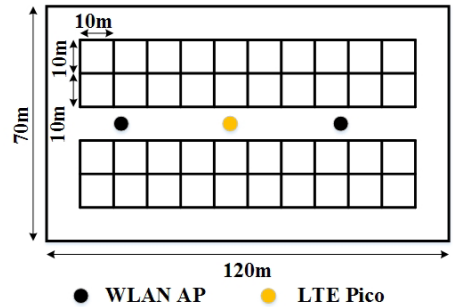}
\vspace{-10pt}
\caption{Simulation scenario.}\label{fig:1}
\vspace{-6pt}
\end{figure}

Considering that WLAN hot-spots are mainly deployed indoors, here we choose a single floor dual-strip model \cite{10} for indoor scenario. This model contains two rows of buildings with 20 rooms in a row and a 10 m wide corridor, as illustrated in Fig. \ref{fig:1}.
\par

In the simulations, 1 Pico and 2 APs are randomly deployed in the corridor with the minimum distance of 10 m. Both LTE and WLAN users are created with a uniform spatial distribution all over the floor. Users may reside either in the rooms or in the corridors and they are stationary during the simulations. There are three Pico/APs in simulations which use the same frequency channel. Main simulation parameters are listed in Table \ref{table:2}. The working frequency band is 5 GHz here. Besides, we assume that LTE has overlapping downlink (DL) bandwidths with WLAN and the user traffic is DL only.
\par
\begin{table}[t]
\caption{Main Deployment Parameters.}
\label{table:2}
\centering
\begin{tabular}{l|l}
\hline
\textbf{Parameters} & \textbf{Value}  \\
\hline
Scenario & Indoor dual-stripe model \\
System Bandwidth & 20.0 MHz \\
Center Frequency & 5.0 GHz \\
Transmission Power & 23.0 dBm \\
Antenna Gain after Cable Loss & 3.0 dBm \\
Traffic Direction & Downlink \\
Traffic & Full-Buffer \\
Call Arrival & Poisson Distributed \\
Signal Propagation & Indoor Pathloss Model \cite{11} \\
Number of Tx/Rx Antennas & 1/1 \\
Antenna Pattern & Omni-Directional\\
Pico/AP Height & 3.0 m \\
UE Height & 1.5 m \\
\hline
\end{tabular}
\vspace{-15pt}
\end{table}
\par
LTE frame is divided into 10 subframes with 1 ms duration. The modulation and coding scheme (MCS) is chosen based on the signal-to-interference-plus-noise ratio (SINR) measured 2 ms ago at each node. Chase combining (CC) is employed for successful hybrid automatic repeat request (HARQ) retransmissions.
\par
The distributed coordination function (DCF) protocol based on CSMA/CA mechanism is adopted in simulations. A markov state machine is utilized to simulate this mechanism of WLAN where two states, backoff level and backoff counter, are included. The size of WLAN backoff window (backoff counter) is defined by backoff level. Backoff counter decreases by one as long as the channel detected by WLAN is idle for a backoff time slot. However, backoff counter will be frozen when the channel is sensed busy and resumed when the channel is idle again. WLAN nodes will start transmitting when backoff counter returns to 0.

\subsection{Simulation Results}

\subsubsection{Adaptive Co-existence Mechanism Simulation Results}
\par
In this work, we simulate performances of LTE and WLAN co-existing in the same frequency band in different LTE traffic arrival rate. LTE traffic arrival rate $\lambda_L$ belongs to $\{0.5,1,1.5,2\}$ per ms. When fixing $\lambda_L$, WLAN traffic arrival rate $\lambda_W$ increases from 0.01 per ms to 1.5 per ms linearly. This is because WLAN traffic load changes along time in reality and only in this way can our adaptive co-existence mechanism work effectively.
\par
\begin{figure}[t]
\centering
\includegraphics[width=0.48\textwidth]{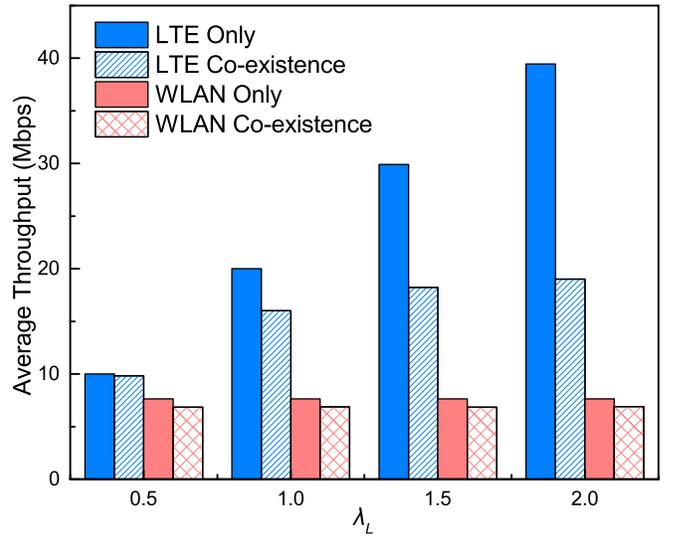}
\vspace{-5pt}
\caption{Throughput results for 1 Pico/2 APs in a single floor dual-stripe scenario. LTE and WLAN co-existing under adaptive co-existence mechanism. The optimum performances of LTE and WLAN are presented as well.}
\vspace{-20pt}
\label{fig:2}
\end{figure}
\par
Three groups of simulations are depicted in Fig. \ref{fig:2}, where \emph{LTE Only} means that only LTE pico occupies the frequency band and works in different $\lambda_L$, \emph{WLAN Only} means that only WLAN APs occupy the frequency band and \emph{LTE Co-existence} $\&$ \emph{WLAN Co-existence} present the performances of LTE and WLAN when they share the same frequency band under different $\lambda_L$. Since $\lambda_W$ increases from 0.01 per ms to 1.5 per ms in every simulation, it is reasonable that the performance of \emph{WLAN Only} remains the same under different $\lambda_L$. Observing \emph{LTE Only} in Fig. \ref{fig:2}, when $\lambda_L$ multiplies in simulations, LTE average throughput multiplies correspondingly. However, when LTE co-exists with WLAN in adaptive co-existence mechanism, its throughput decreases inevitably since LTE spares subframes to WLAN. \emph{LTE Co-existence} performances lose 1.92\%, 19.97\%, 39.06\% and 51.82\% compared to the optimum performances of \emph{LTE Only} when $\lambda_L$ is 0.5 per ms, 1 per ms, 1.5 per ms and 2 per ms, respectively. On the other hand, WLAN performs significantly compared to its optimum performance when it co-exists with LTE in adaptive co-existence mechanism. \emph{WLAN Co-existence} performances lose 10.39\%, 10.00\%, 10.26\% and 9.90\% compared to \emph{WLAN Only} performances in different $\lambda_L$, respectively. The adaptive co-existence mechanism ensures WLAN system performs almost as good as \emph{WLAN Only} when it co-exists with LTE in the same frequency band.
\par
\begin{figure}[t]
\centering
\includegraphics[width=0.48\textwidth]{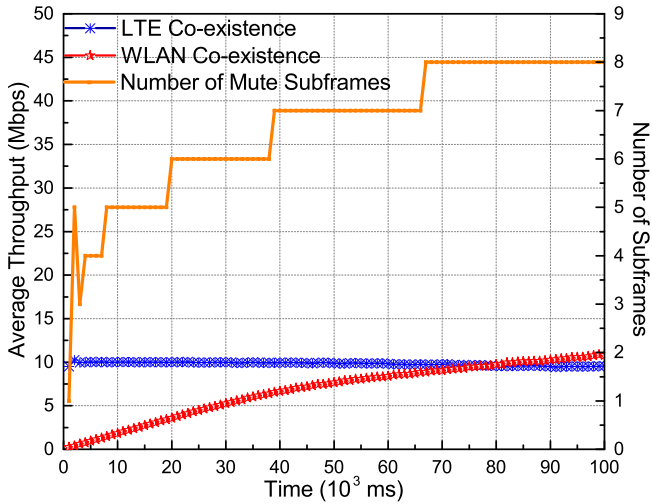}
%\vspace{-5pt}
\caption{Collections of LTE and WLAN average throughput and the number of subframes given to WLAN during simulations, $\lambda_L$ = 0.5 per ms.}
\label{fig:3}
%\vspace{-5pt}
\end{figure}
\begin{figure}[t]
\centering
\includegraphics[width=0.48\textwidth]{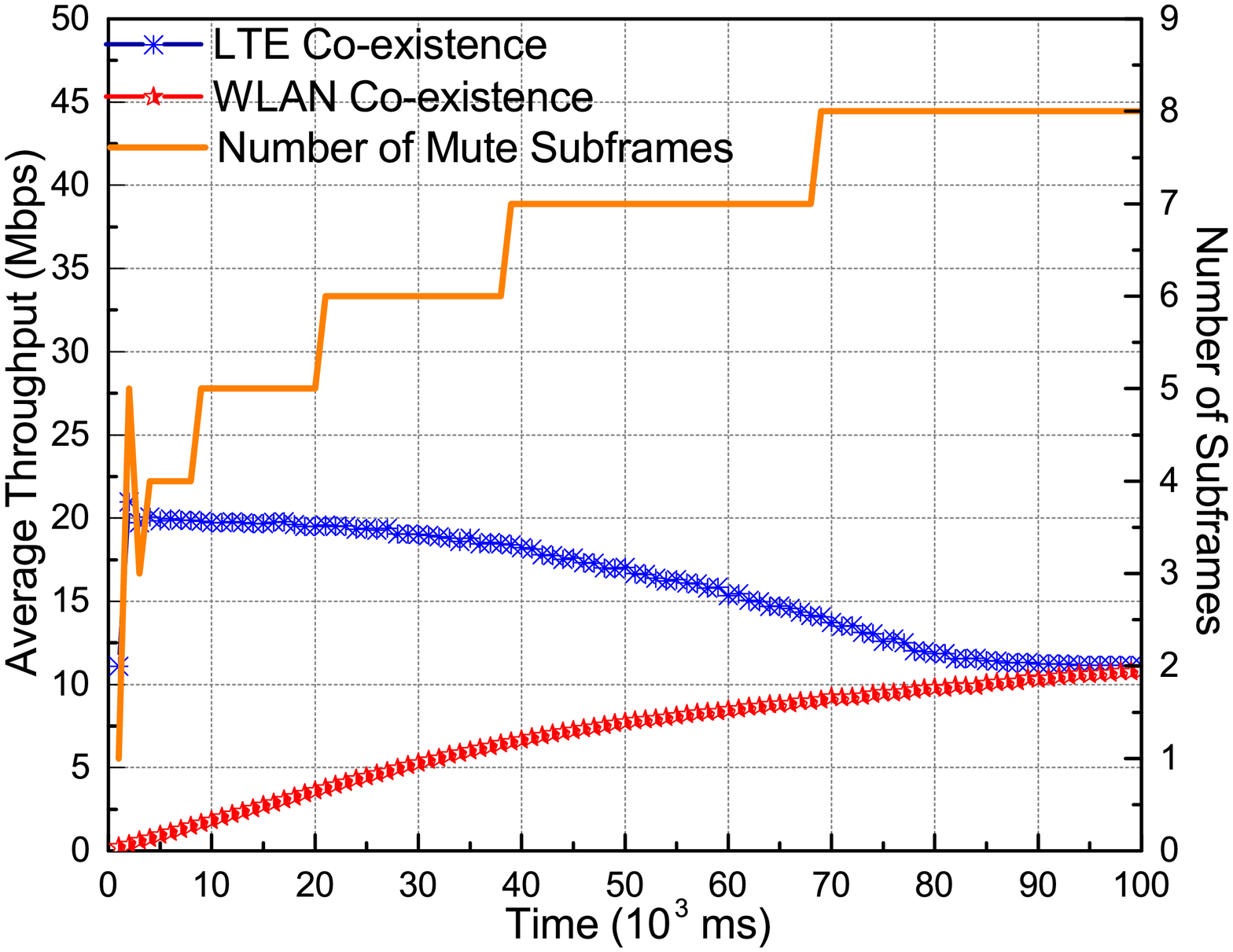}
%\vspace{-5pt}
\caption{Collections of LTE and WLAN average throughput and the number of subframes given to WLAN during simulations, $\lambda_L$ = 1 per ms.}
\label{fig:4}
%\vspace{-15pt}
\end{figure}
\begin{figure}[t]
\centering
\includegraphics[width=0.48\textwidth]{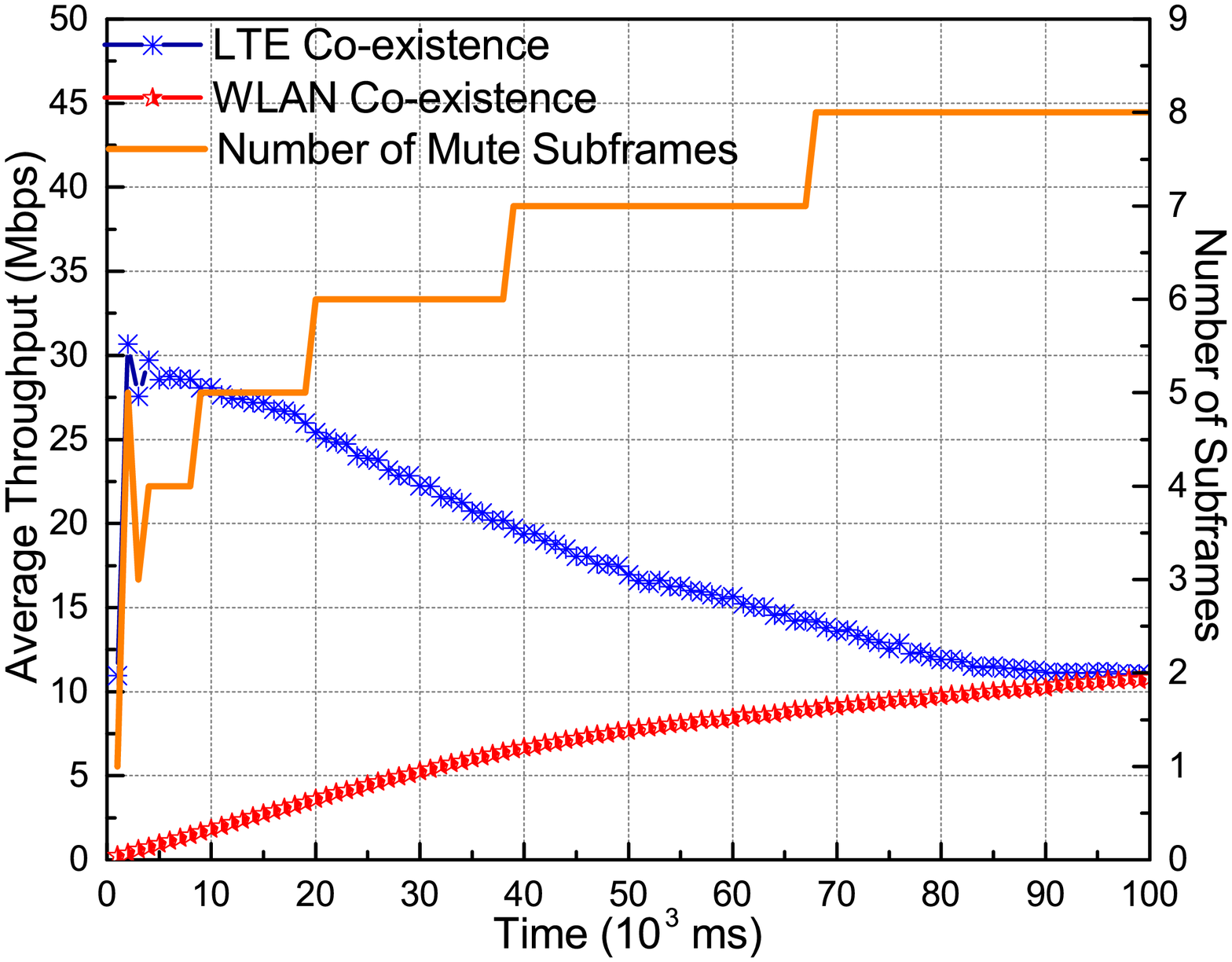}
%\vspace{-5pt}
\caption{Collections of LTE and WLAN average throughput and the number of subframes given to WLAN during simulations, $\lambda_L$ = 1.5 per ms.}
\label{fig:5}
%\vspace{-5pt}
\end{figure}
\begin{figure}[t]
\centering
\includegraphics[width=0.48\textwidth]{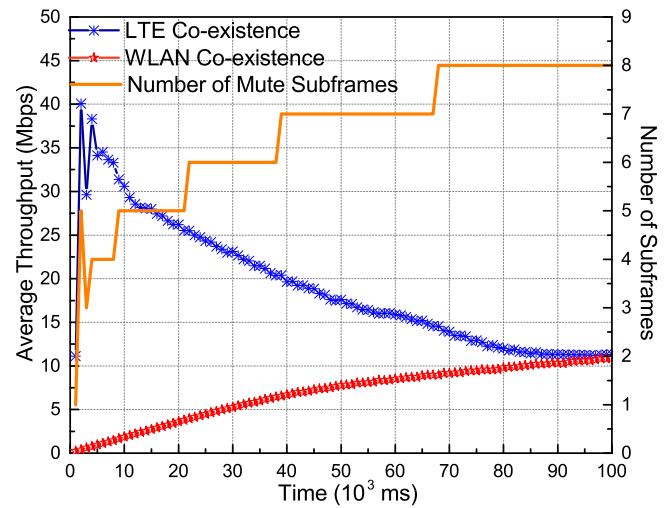}
%\vspace{-5pt}
\caption{Collections of LTE and WLAN average throughput and the number of subframes given to WLAN during simulations, $\lambda_L$ = 2 per ms.}
\label{fig:6}
%\vspace{-15pt}
\end{figure}

In the co-existence scenario, we also evaluate average throughput of LTE and WLAN and the number of subframes spared to WLAN system every $T_c$. The total simulation time is 100,000 ms and $T_c$ is 1,000 ms so that we collect 100 average throughput in one simulation drop. We simulate 500 drops and calculate the average performances of both systems, and the simulation results are presented in Figs. \ref{fig:3}-\ref{fig:6}.

\par
From these figures we observe that LTE can track WLAN traffic load and adjust the number of mute subframes efficiently. WLAN traffic arrival rate increases gradually over time so that the throughput of WLAN increases correspondingly, which indicates that it suffers little from the LTE interference. This is because LTE gives subframes to WLAN according to its traffic load. Fig. \ref{fig:3} shows that when $\lambda_L$ is small, LTE can finish most of the transmission even though it spares most subframes. In this simulation, the average throughput of LTE remains 10 Mbps. Meanwhile, $\lambda_L$ is 0.5 per ms and packet size of LTE is 20 kbits, so the traffic load of LTE in Fig. \ref{fig:3} is 10 Mbps, as well. Therefore, co-existing with WLAN causes little throughput loss on LTE system. However, when traffic load of LTE is heavier in Figs. \ref{fig:5} and \ref{fig:6}, LTE can not finish transmission when it still gives the majority of subframes to WLAN. As a result, its average throughput decreases severely when the number of mute subframes increases gradually.

\subsubsection{Comparison of Adaptive Co-existence Mechanism with Fixed Mute Subframes Co-existence Mechanism}
\par
In this part, we compare the system performances of LTE co-existing with WLAN in adaptive co-existence mechanism and the fixed subframe configuration mechanisms. Tables \ref{table:3} and \ref{table:4} list the simulated throughput of LTE and WLAN, respectively.
\begin{table}[t]
\caption{WLAN Specific Values of Average Throughput.}
\vspace{-5pt}
\label{table:3}
\centering
\begin{tabular}{l|l|l|l|l|l|l}
\hline
 & \textbf{Best} &\textbf{Adaptive} &\textbf{Mode1} & \textbf{Mode2} &\textbf{Mode3} &\textbf{Mode4}\\
\hline $\lambda_{L_{0.5}}$ & 7.641 & 6.847 & 4.047 & 5.432 & 6.356 & 7.124 \\
\hline $\lambda_{L_1}$ & 7.641& 6.877& 4.079& 5.425& 6.357& 7.134\\
\hline $\lambda_{L_{1.5}}$ & 7.641& 6.857& 4.009& 5.449& 6.376& 7.072\\
\hline $\lambda_{L_2}$ & 7.641& 6.885& 4.063& 5.420& 6.366& 7.148\\
\hline
\end{tabular}
\vspace{-5pt}
\end{table}

\begin{table}[t]
\caption{LTE Specific Values of Average Throughput.}
\vspace{-5pt}
\label{table:4}
\centering
\begin{tabular}{l|l|l|l|l|l|l}
\hline
 & \textbf{Best} &\textbf{Adaptive} &\textbf{Mode1} & \textbf{Mode2} &\textbf{Mode3} &\textbf{Mode4}\\
\hline $\lambda_{L_{0.5}}$ & 10.006 & 9.814 &10.007 &10.004 & 9.992 & 9.636 \\
\hline $\lambda_{L_1}$ & 20.008& 16.012&19.992&19.905& 19.426& 11.272\\
\hline $\lambda_{L_{1.5}}$ & 29.897& 18.220& 29.673& 28.971& 22.642& 10.998\\
\hline $\lambda_{L_2}$ & 39.440& 19.003& 38.732& 34.208& 22.607& 11.225\\
\hline
\end{tabular}
\vspace{-15pt}
\end{table}
\par
\begin{figure}[ht]
\vspace{-10pt}
\centering
\includegraphics[width=0.48\textwidth]{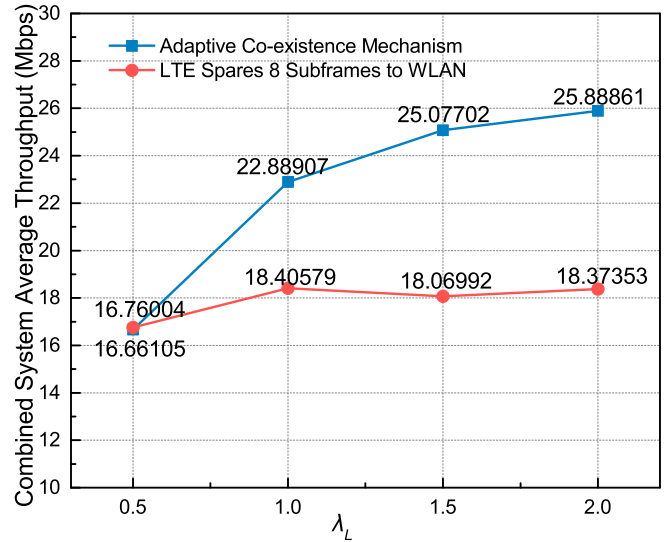}
\vspace{-5pt}
\caption{Combined throughput of LTE and WLAN co-existing in adaptive co-existence mechanism and in Mode 4.}
\label{fig:7}
\vspace{-5pt}
\end{figure}

From Table \ref{table:3}, if we guarantee that the throughput loss of WLAN system is less than 15\% in co-existence, LTE can only work in Mode 4 and adaptive co-existence mechanism. From Table \ref{table:3} we can see that in Mode 4, WLAN performance is a little better than that of adaptive co-existence mechanism. However, from Table \ref{table:4}, it is obvious that LTE with adaptive co-existence mechanism performs much better than that in Mode 4. This is because LTE allocates less than 8 subframes to WLAN system during most of the simulation time in adaptive co-existence mechanism. The number of mute subframes is adjusted adaptively. On the other hand, LTE always gives 8 subframes to WLAN system in Mode 4, which leads to higher performance of WLAN system while the performance of LTE system degrades a lot. The combined throughput of LTE and WLAN in these mechanisms are presented in Fig. \ref{fig:7}.
\par
Fig. \ref{fig:7} shows that the adaptive co-existence mechanism outperforms Mode 4 in terms of combined average throughput of LTE and WLAN. In simulations, WLAN traffic load increases linearly while in reality, it will change randomly, which indicates that the combined performance of LTE and WLAN may be better than other mechanisms when the average throughput of WLAN is expected to lose 15\% at most. From the simulation results we have presented, the adaptive co-existence mechanism will bring much better performance to WLAN.
\section{Conclusion}
In this paper, an adaptive co-existence mechanism of LTE co-existing with WLAN in the same unlicensed band is proposed. Its performance is verified to outperform the existing ones. If LTE works in a fixed subframe configuration mode, the throughput of WLAN will increase gradually while that of LTE will decrease at the same time. From simulations, we have found that the adaptive co-existence mechanism guarantees a significant performance of WLAN system which only loses about 10\% compared to the WLAN optimum performance. Collections of LTE and WLAN average throughput and number of mute subframes every reallocation cycle in simulations prove that LTE can track WLAN traffic load and modify its subframe configurations correspondingly. As a result, the adaptive co-existence mechanism brings a better combined throughput when WLAN loses no more than 15\% average throughput. Future research is needed to find more sophisticated co-existence mechanisms that can ensure higher LTE performance when it co-exists with WLAN while WLAN system performance is also significant.

\section*{Acknowledgment}

This work is supported in part by the Fundamental Research Funds for the Central Universities (No. 2014ZD03-02), the National Key Technology R$\&$D Program of China (No. 2014ZX03004002) and the National Natural Science Foundation of China under Grant 61171106.

\end{document}